\documentclass[aps, pra, reprint, longbibliography, onecolumn, notitlepage]{revtex4-1}

\usepackage[utf8]{inputenc}
\usepackage[english]{babel}
\usepackage[colorlinks, citecolor=blue]{hyperref}
\usepackage{graphicx}
\usepackage{amssymb, amsmath}
\usepackage{bm}
\usepackage{siunitx}

\begin{document}

\title{Stationary Gaussian entanglement between levitated nanoparticles}

\author{Anil Kumar Chauhan}
\email{anil.kumar@upol.cz}
\affiliation{Department of Optics, Palack\'y University, 17. listopadu 1192/12, 77146 Olomouc, Czechia}

\author{Ond\v{r}ej \v{C}ernot\'{i}k}
\email{ondrej.cernotik@upol.cz}
\affiliation{Department of Optics, Palack\'y University, 17. listopadu 1192/12, 77146 Olomouc, Czechia}

\author{Radim Filip}
\email{filip@optics.upol.cz}
\affiliation{Department of Optics, Palack\'y University, 17. listopadu 1192/12, 77146 Olomouc, Czechia}

\date{\today}

\begin{abstract}
	Coherent scattering of photons is a novel mechanism of optomechanical coupling for optically levitated nanoparticles promising strong, versatile interactions with light and between nanoparticles.
	We show that it allows efficient deterministic generation of Gaussian entanglement between two particles in separate tweezers.
	A combination of red- and blue-detuned tweezers brings a mechanical Bogoliubov mode to its ground state.
	An additional, dispersively coupled cavity mode can reduce noise in the orthogonal mode, resulting in strong entanglement as quantified by the logarithmic negativity and verifiable with the Duan criterion for realistic experimental parameters.
	Such an important resource for quantum sensing and quantum simulations is pivotal for current experiments and presents an important step towards optomechanics with multiple particles in the quantum regime.
\end{abstract}

\maketitle

\section{Introduction}
Nonclassical states of macroscopic mechanical resonators offer a unique opportunity to investigate the boundary between classical and quantum worlds~\cite{Romero-Isart2011a,Arndt2014,Bateman2014,Pino2018,Pontin2019}.
Particularly mechanical modes formed by center-of-mass motion of optically levitated particles~\cite{Romero-Isart2011,Chang2010,Millen2020} attracted attention in this regard as they do not suffer from clamping losses, allowing for efficient isolation from thermal noise. Moreover, optical trapping allows versatile modification of the potential in space and time~\cite{fonseca2016,Siler2018,tebbenjohanns2019}.
Over the years, the field of levitated optomechanics reached remarkable level of control over the degrees of freedom down to the quantum regime~\cite{Delic2020,Tebbenjohanns2020}, using optomechanical interactions not only for investigating the quantum-to-classical transition but also for  thermodynamics~\cite{Dechant2015,Gieseler2018,Debiossac2020}, force sensing~\cite{Moore2014,Rider2016,Blakemore2019,Monteiro2020}, or interactions with other quantum systems~\cite{Nie2016,Chen2017}.
Recently, coherent scattering of tweezer photons into an empty cavity mode arose as an interesting coupling mechanism~\cite{Gonzalez-Ballestero2019}, allowing efficient cooling~\cite{Delic2019,Windey2019} (even down to the quantum ground state~\cite{Delic2020}) and strong optomechanical coupling~\cite{sommer2020strong}. Theoretical proposals suggest to employ it for three-dimensional displacement detection~\cite{Toros2019} or for preparation of nonclassical mechanical states~\cite{Cernotik2020,Rudolph2020}.

As the level of control of levitated particles progresses, systems involving multiple mechanical and electromagnetic modes become attractive for quantum sensing, simulations, and thermodynamics.
These advances follow the development of cavity optomechanics with clamped mechanical resonators in which theoretical proposals and experimental demonstrations addressed generation of nonclassical correlations between fields~\cite{Tian2013,Wang2013,Barzanjeh2019,Chen2020} or between photons and phonons~\cite{Mari2009,Palomaki2013,Riedinger2016,Gut2019,Brunelli2020,Rakhubovsky2020}.
Entanglement between mechanical modes can also be prepared, for example, by measurements~\cite{Borkje2011,Riedinger2018}, via parametric or blue-sideband driving~\cite{Hofer2015,Pontin2016,Kotler2020,Lin2020}, or using reservoir engineering~\cite{Tan2013,Woolley2014,Li2015,Ockeloen-Korppi2018}.
Particularly the last strategy is attractive as it allows Gaussian entanglement to be generated deterministically in the steady state.
Proposals for entanglement generation tailored for levitated systems exist as well but they either rely on weak nonlinear interactions~\cite{Abdi2015} or employ photon counting~\cite{Rudolph2020} and are highly susceptible to thermal noise.

\begin{figure}
\includegraphics[width=0.6\linewidth]{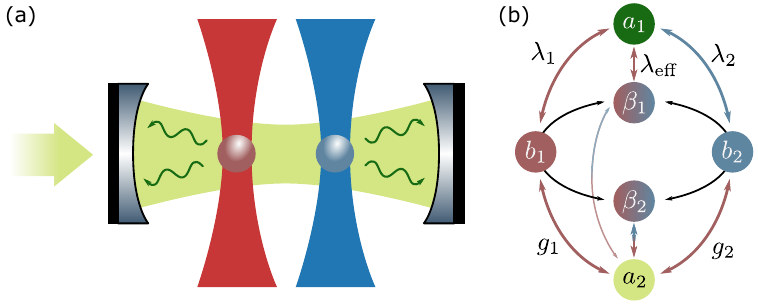}
\caption{\label{sys}
 	Schematic of the proposed setup.
 	(a) Two levitated particles are trapped in two tweezers of different frequencies and coupled to a cavity mode via coherent scattering ($a_1$, dark green);
 	another cavity mode ($a_2$, light green) is driven by an external pump and coupled to the particle motion dispersively.
 	(b) The two mechanical modes $b_{1,2}$ (coupled to the cavity mode $a_1$ via beam-splitter (red arrow) and two-mode squeezing (blue) interaction, respectively) form two Bogoliubov modes $\beta_{1,2}$ which are predominantly coupled to one cavity mode each.
 	The first Bogoliubov mode $\beta_{1}$ couples via coherent scattering to the cavity mode $a_1$ with only a weak coupling to the second cavity mode $a_{2}$.
 	The second Bogoliubov mode $\beta_{2}$ interacts only with $a_2$.}
\end{figure}

Here, we propose and analyse a scheme to create Gaussian two-mode squeezed entangled states of two nanoparticles deterministically in the steady state.
The particles are levitated by two separate tweezers and coupled to the same optical cavity mode via coherent scattering.
A suitable combination of tweezer detunings (the first tweezer is detuned to the red sideband while the second tweezer is detuned to the blue sideband; see figure~\ref{sys}) allows us to cool a collective mechanical Bogoliubov mode to its quantum ground state~\cite{Tan2013,Woolley2014,Li2015}.
To combat the strong thermal noise present in the orthogonal Bogoliubov mode and open a road to entanglement generation in realistic conditions, we employ a second cavity mode coupled to both particles via the more common dispersive coupling and driven on the red sideband by an external laser~\cite{Aspelmeyer2014}. 
Unlike existing proposals for dissipative preparation of two-mode squeezing (which require up to four driving fields~\cite{Woolley2014,Ockeloen-Korppi2018}), our scheme requires only one external drive and can generate nonclassical correlations between two mechanical modes starting from room temperature in state-of-the-art levitated systems.
This not only significantly reduces the experimental complexity but also drastically curtails the effects of laser phase noise associated with the classical laser drives.
Our work thus presents an important step towards exploring quantum effects in the collective motion of multiple levitated particles and to the applications mentioned above.

\section{Dissipative generation of phonon–phonon entanglement}

\subsection{Cooling a mechanical Bogoliubov mode via coherent scattering}

We consider two particles trapped in separate tweezers and placed inside a cavity as shown in figure~\ref{sys}(a).
For each particle, the tweezer defines a harmonic potential with frequency $\Omega_j$, $j=1,2$, and scattering of the tweezer photons into the cavity provides the optomechanical interaction.
As described by Gonzalez-Ballestero \emph{et al.}~\cite{Gonzalez-Ballestero2019}, the interaction includes the trapping potential of the tweezers (given by $-\frac{1}{2}\alpha E_j^2(x_j)$, where $E_j(x_j)$ is the electric field of the $j$th tweezer at the position of the particle $x_j$), the usual dispersive optomechanical interaction (given by a similar expression for the electric field of the cavity mode with annihilation operator $a_1$) of the form $G_j a_1^\dagger a_1 x_j$ (which is much weaker than coherent scattering and we neglect it in the following), and the coupling mediated by coherent scattering (given by the product of the tweezer and cavity fields) given by $\lambda_j(a_1+a_1^\dagger)(b_j+b_j^\dagger)(e^{i\nu_j t} + e^{-i\nu_j t})$, where $\nu_j$ is the frequency of the $j$th tweezer.

Moving to a frame in which the cavity mode $a_1$ oscillates at the frequency of the first tweezer, $\nu_1$, we have
\begin{equation}\label{rfHamil}
	H = \Delta_{1} a_{1}^\dag a_{1}+ \Omega_1b_1^\dagger b_1+\Omega_2 b_2^\dagger b_2 
	-\lambda_{1}(a_{1}+a_{1}^\dag)(b_1+b_1^\dagger)
	-\lambda_{2}(a_{1} e^{i\delta t}+a_1^\dagger e^{-i\delta t})(b_2+b_2^\dagger),
\end{equation}
where $\Delta_1 = \omega_1 - \nu_1$ is the detuning between the first tweezer and the cavity resonance $\omega_1$ and $\delta = \nu_2-\nu_1$ is the detuning between the two tweezers.
Next, we set the detunings as $\Delta_1 = \Omega_1$, $\delta = \Omega_1 + \Omega_2$ and move to the rotating frame with respect to the free oscillations of all three modes, $\Delta_1 a_1^\dagger a_1 + \Omega_1 b_1^\dagger b_1 + \Omega_2 b_2^\dagger b_2$;
under the rotating wave approximation, we obtain the coherent-scattering Hamiltonian
\begin{equation}\label{int_Ham}
	H_{\rm cs}=-\lambda_1(a_{1}^\dagger b_{1}+b_{1}^\dagger a_{1})-\lambda_2(a_{1}b_{2}+ a_{1}^\dagger b_{2}^\dagger).
\end{equation}
To see how this Hamiltonian leads to steady-state two-mode squeezing, we introduce the Bogoliubov mode $\beta_1 = (\lambda_1 b_1+\lambda_2 b_2^\dagger)/\lambda_{\rm eff}$, where $\lambda_{\rm eff} = \sqrt{\lambda_1^2-\lambda_2^2}$~\cite{Woolley2014}.
The Hamiltonian \eqref{int_Ham} then becomes $H_{\rm cs} = -\lambda_{\rm eff}(a_1^\dagger\beta_1+\beta_1^\dagger a_1)$;
this interaction cools the Bogoliubov mode $\beta_1$ to its ground state, producing two-mode squeezing between the original modes $b_j$ (see also figure~\ref{sys}(b)).

The dynamics of this system (in terms of the particle modes $b_j$) are described by the Langevin equations
\begin{subequations}
\begin{align}\label{Int_eqs}
\dot{a}_{1} &= i(\lambda_{1}b_{1}+\lambda_{2}b_{2}^\dagger) -\frac{\kappa_{1}}{2}a_{1}+\sqrt{\kappa_{1}}a_{1, {\rm in}}, \\
\dot{b}_{1} &= i\lambda_{1}a_{1}-\frac{\gamma_1}{2}b_{1}+\sqrt{\gamma_{1}}b_{1,{\rm in}}, \\
\dot{b}_{2} &= i\lambda_{2}a_{1}^\dagger-\frac{\gamma_2}{2}b_{2}+\sqrt{\gamma_2} b_{2,{\rm in}},
\end{align}
\end{subequations}
where $\kappa_{1}$ is the decay rate of the cavity mode $a_{1}$ and $\gamma_{j}$ are the damping rates for the two mechanical modes. The input noise $a_{1,{\rm in}}$ has the correlations $\langle a_{1,{\rm in}}(t)a_{1,{\rm in}}^\dagger(t')\rangle=\delta(t-t')$; the thermal noise of the mechanical modes has the correlation function $\langle b_{j,{\rm in}}(t)b_{k,{\rm in}}^\dagger(t')\rangle = (2n_{j}+1)\delta_{jk}\delta(t-t')$, where $n_{j}\simeq k_BT/\hbar\Omega_j$ is the mean thermal occupation for mechanical frequency $\Omega_j$ at temperature $T$.
In terms of the cavity quadrature operators $X_1 = (a_1+a_1^\dagger)/\sqrt{2}, Y_1 = -i(a_1-a_1^\dagger)/\sqrt{2}$ and canonical mechanical operators $x_j = (b_j+b_j^\dagger)/\sqrt{2}, p_j = -i(b_j-b_j^\dagger)/\sqrt{2}$, the dynamics can be expressed in the form
\begin{subequations}
\begin{align}
	\dot{{r}} &= {Ar}+{\xi},\\
	{r} &= (X_1,Y_1,x_1,p_1,x_2,p_2)^T,\\
	{\xi} &= (\sqrt{\kappa_1}X_{1,{\rm in}},\sqrt{\kappa_1}Y_{1,{\rm in}}, \sqrt{\gamma_1}x_{1,{\rm in}},\sqrt{\gamma_1}p_{1,{\rm in}}, \sqrt{\gamma_2}x_{2,{\rm in}},\sqrt{\gamma_2}p_{2,{\rm in}})^T \label{eq:SMnoise}\\
	{A}& = \begin{pmatrix}
		-\frac{1}{2}\kappa_{1} &0& 0& -\lambda_{1} & 0& \lambda_{2}\\
		0& -\frac{1}{2}\kappa_{1} & \lambda_{1} & 0& \lambda_{2}  & 0\\
		0 & -\lambda_{1} & -\frac{1}{2}\gamma_{{1}} & 0 &   0 & 0 \\
		\lambda_{1} & 0& 0&  -\frac{1}{2}\gamma_{{1}}& 0 & 0 \\
		0&\lambda_{2} & 0 &0 & -\frac{1}{2}\gamma_{{2}} & 0\\
		\lambda_{2} & 0 & 0 &0& 0 & -\frac{1}{2}\gamma_{2}
	\end{pmatrix},
\end{align}
\end{subequations}
where the input noise quadrature operators are defined in full analogy to the quadrature operators of the cavity and mechanical modes.

To evaluate the generated entanglement, we solve for the steady-state covariance matrix associated with these dynamical equations as described in the appendix.
From the mechanical covariance matrix $V$, we then calculate the logarithmic negativity $E_n$  which is an entanglement measure for Gaussian states~\cite{Vidal2002} and the state purity $P = 1/\sqrt{\det V}$.
For a direct experimental validation of the generated entanglement, we also consider the violation of the Duan criterion~\cite{Duan2000}
\begin{equation}\label{eq:EPR}
	\Delta_{\rm EPR}=\frac{1}{2}[\Delta( x_{1}+x_{2})+\Delta(p_{1}-p_{2})]\geq 1,
\end{equation}
where $\Delta (O)=\langle O^2\rangle-\langle O\rangle^2$ is the variance of the operator~$O$.
The violation of the Duan criterion (which is a sufficient condition for entanglement) is easier to verify experimentally as it involves only a pair of commuting operators to be measured whereas determination of the logarithmic negativity requires full tomography to obtain the whole covariance matrix $V$.
To further highlight the difference between our results and previous work~\cite{Rudolph2020}, we also calculate the noise reduction factor~\cite{Iskhakov2009} which is defined as the variance of the phonon-number difference normalised by the total number of excitations,
\begin{equation} \label{deln}
{\rm NRF}=\frac{\langle \left(n_{1}-n_{2} \right)^2\rangle-\langle \left(n_{1}-n_{2}\right)\rangle^2}{\langle n_1+n_2\rangle},
\end{equation}
 where the phonon number $n_{j}=b_{j}^\dag b_{j}$.
 Noise reduction factor smaller than unity is a clear signature of sub-Poissonian phonon--phonon correlations (and it vanishes completely for pure two-mode squeezing).
 Conversely, single-phonon entangled states proposed in~\cite{Rudolph2020} exhibit phonon anti-correlation between particles with ${\rm NRF} = 1$.

\begin{figure}
	\centering
	\includegraphics[width=\linewidth]{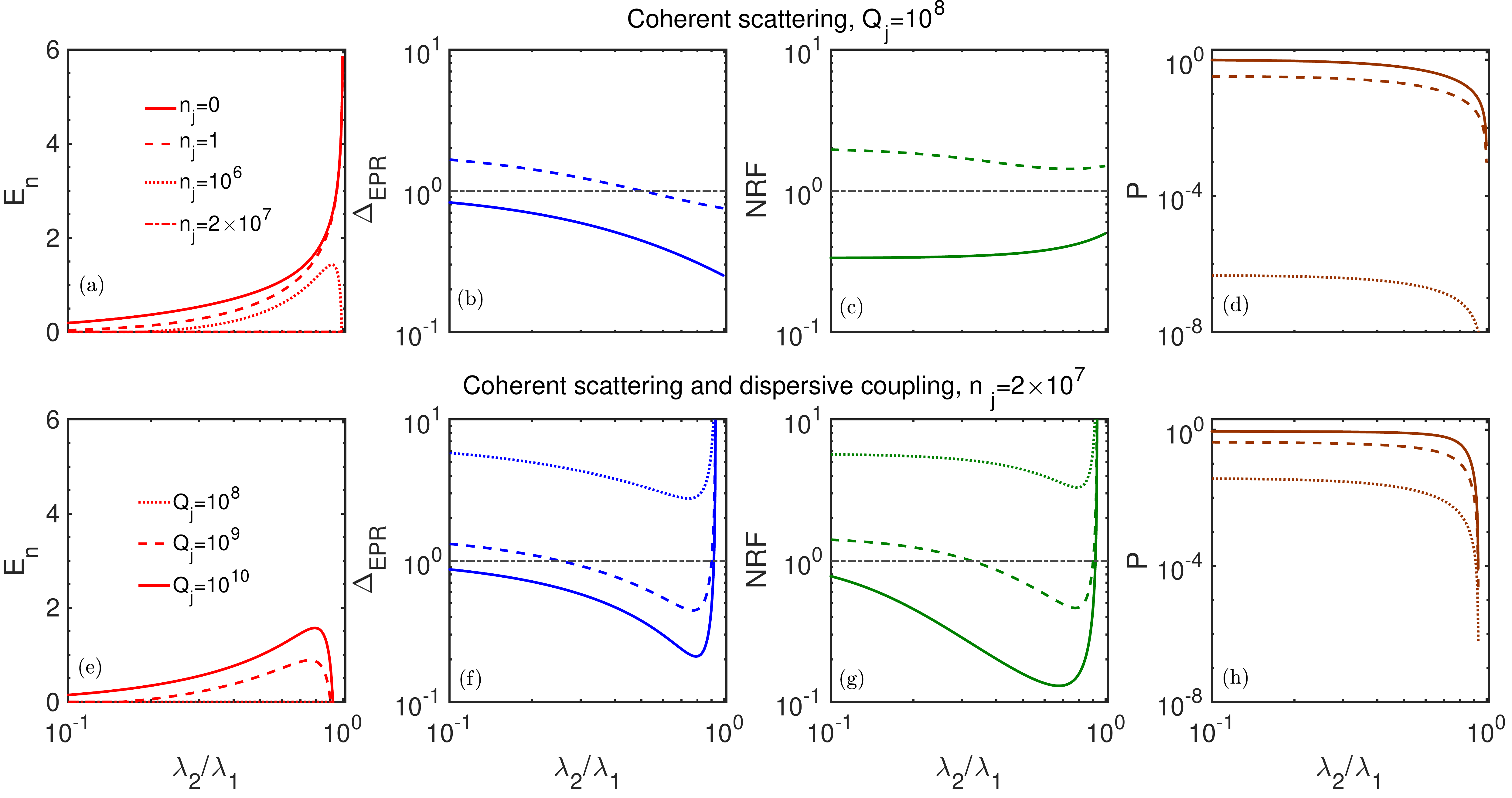}
	\caption{\label{fig:results}
		(a)--(d) Generation of two-particle entanglement using coherent scattering with the Hamiltonian $H_{\rm cs}$.
		We plot the logarithmic negativity (a), EPR variance (b), noise reduction factor (c), and mechanical-state purity (d) versus the ratio of the coupling strengths $\lambda_2/\lambda_1$ for the thermal noise levels $n_j=0$ (solid lines), $n_j=1$ (dashed), $n_j=10^6$ (dotted), and $n_j=2\times 10^7$ (dot-dashed) with the mechanical quality factor $Q_j=\Omega_j/\gamma_j=10^8$.
		(e)--(h) Entanglement generation using a combination of coherent scattering and dispersive optomechanics as described by the Hamiltonian $H = H_{\rm cs}+H_{\rm disp}$;
		here, we consider thermal noise $n_j=2\times 10^7$ and mechanical quality factors $Q_j=10^8$ (dotted lines), $Q_j=10^9$ (dashed), and $Q_j=10^{10}$ (solid).
		The remaining parameters, equal for both sets of plots, are $\lambda_1/2\pi=\SI{100}{\kilo\hertz}$ and $\kappa_1/2\pi = \SI{120}{\kilo\hertz}$.
		For panels (e)--(h), the dispersive coupling strengths are $g_1/2\pi = \SI{3}{\kilo\hertz}$ and $g_2/2\pi = \SI{20}{\kilo\hertz}$ and we assume equal mechanical frequencies, $\Omega_1 /2\pi= \Omega_2/2\pi= \SI{300}{\kilo\hertz}$.
		In panels (b), (f), the horizontal black line gives the classical limit;
		EPR variance smaller than this bound, $\Delta_{\rm EPR}<1$, implies that the two particles are entangled.
		In panels (c), (g), noise reduction factor below the horizontal line at unity signifies sub-Poissonian phonon-number correlations between the two mechanical modes.}
\end{figure}

The results of these simulations are shown in figure~\ref{fig:results}(a)--(d).
For these plots, we consider experimental parameters similar to the recent coherent scattering experiments~\cite{Delic2019,Delic2020}.
Entanglement between the two particles can be generated for thermal noise levels surpassing $n_j=10^6$ as verified by the logarithmic negativity (panel (a));
for mechanical frequency of \SI{305}{\kilo\hertz}, this noise level corresponds to a temperature of about \SI{15}{\kelvin}.
With high levels of thermal noise, the logarithmic negativity drops to zero for $\lambda_2\to\lambda_1$ as the coupling rate of the Bologiubov mode $\lambda_{\rm eff}\to 0$.
The EPR variance, on the other hand, drops below the classical limit only for very small amounts of thermal noise, $n_j\lesssim 1$ (panel (b)); for $n_j=10^6$ (not shown), the minimum attainable EPR variance is $\sim 5\times10^5$.
This discrepancy is caused by the inherent asymmetry of the mechanical state where the first mechanical mode $b_1$ is actively cooled by the optomechanical interaction whereas the second mechanical mode $b_2$ is heated up.
The Duan criterion should, in its full generality, still be able to detect entanglement when this asymmetry is taken into account. Such a verification scheme would, however, be more difficult experimentally than a direct measurement of the collective operators $x_1+x_2$ and $p_1-p_2$ and would require full tomography of the covariance matrix.
This asymmetry is further highlighted by the noise reduction factor which remains smaller than unity only for extremely small amounts of thermal noise (panel (c)).
Finally, the state purity quickly drops with increasing thermal noise as shown in panel (d).
These results are understandable since we cool only the Bogoliubov mode $\beta_1$ while the orthogonal Bogoliubov mode $\beta_2 = (\lambda_2 b_1^\dagger + \lambda_1 b_2)/\sqrt{\lambda_{\rm eff}}$ is decoupled from the cavity field and remains in a thermal state.

\subsection{Noise suppression via dispersive coupling}
To make the scheme more noise-tolerant, we use optomechanical coupling to an additional cavity mode $a_2$ which will bring the noise in the mechanical Bogoliubov mode $\beta_2$ down.
As this optical mode will be far detuned from the tweezers, we cannot use coherent scattering to couple the particles to this mode but instead employ the more conventional dispersive optomechanical coupling.
We drive the mode $a_2$ on the red mechanical sideband (here, we assume equal mechanical frequencies, $\Omega_1=\Omega_2$) such that, in the rotating frame and under the rotating wave approximation, the total Hamiltonian becomes $H = H_{\rm cs} + H_{\rm disp}$, where $H_{\rm cs}$ is given in~\eqref{int_Ham} and
\begin{equation}
	H_{\rm{disp}} = g_1(a_{2}^\dagger b_1 + b_1^\dagger a_2) + g_2(a_{2}^\dagger b_2 + b_2^\dagger a_2).
\end{equation}
We can express the dispersive part of the Hamiltonian as $H_{\rm disp} = (g_1\lambda_1\beta_1-g_2\lambda_2\beta_1^\dagger)a_2^\dagger/\lambda_{\rm eff} + (g_2\lambda_1\beta_2-g_1\lambda_2\beta_2^\dagger)a_2^\dagger/\lambda_{\rm eff} + {\rm H.c.}$ which shows that the cavity mode $a_2$ generates single-mode squeezing for both Bogoliubov modes~\cite{Kronwald2013}.
Ideally, the dispersive Hamiltonian $H_{\rm disp}$ should only cool the Bogoliubov mode $\beta_2$ without any additional interactions;
single-mode squeezing introduces asymmetry in the steady state, making the generated entanglement undetectable by the Duan criterion.
For $\lambda_1 > \lambda_2$ (which is necessary for dynamical stability and well-defined Bogoliubov modes) and $g_1 < g_2$, the dominant part of the dispersive Hamiltonian is indeed $g_2\lambda_1(a_2^\dagger\beta_2+\beta_2^\dagger a_2)/\lambda_{\rm eff}$ and the second cavity mode cools the Bogoliubov mode $\beta_2$, reducing the amount of thermal noise in the mechanical steady state.
The small additional squeezing contribution and a weak residual coupling to the first Bogoliubov mode $\beta_1$ slightly alleviate the steady-state noise (see also figure~\ref{sys}(b)).

\begin{figure}
	\centering
	\includegraphics[width=0.95\linewidth]{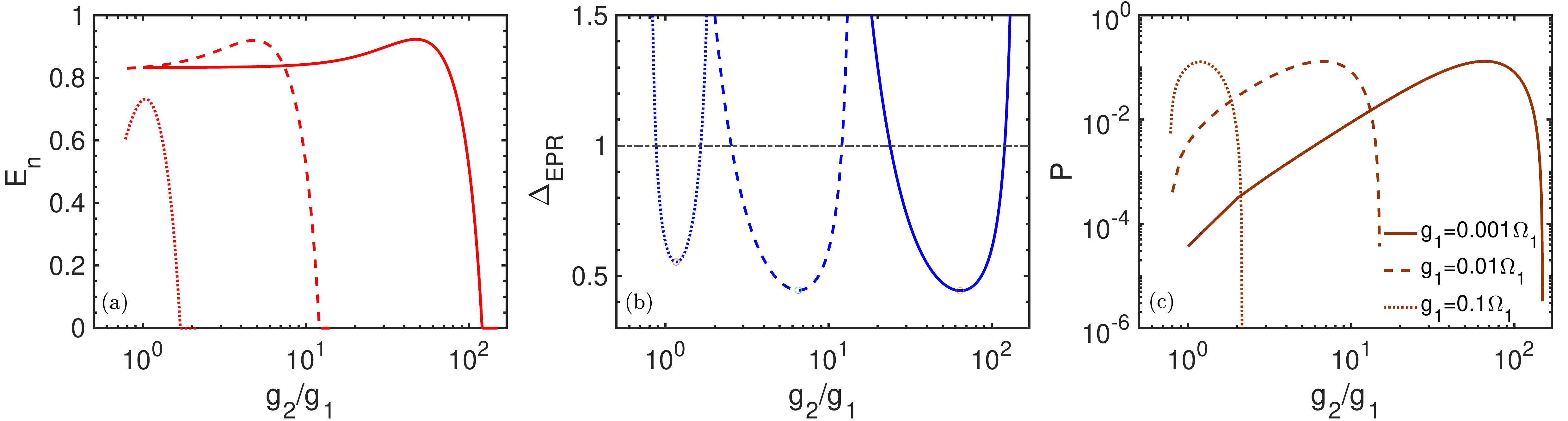}
\caption{\label{fig_3}
	Mechanical steady state as a function of the dispersive coupling.
	The logarithmic negativity (a), EPR variance (b), and state purity (c) are plotted against the ratio of the dispersive couplings $g_2/g_1$ for $g_{1}/2\pi= \SI{300}{\hertz}$ (solid line), $g_{1}/2\pi= \SI{3}{\kilo\hertz}$ (dashed line), and $g_{1}/2\pi=\SI{30}{\kilo\hertz}$ (dotted line).
	The remaining parameters are the same as in figure~\ref{fig:results};
	in addition, we have $\lambda_{2}/2\pi=\SI{80}{\kilo\hertz}$ and $Q_{j}=10^{9}$.
	}
\end{figure}

We plot the resulting entanglement and mechanical state purity in figure~\ref{fig:results}(e)--(h). Efficient generation of entanglement---confirmed by logarithmic negativity, EPR variance, as well as the noise reduction factor---is possible with realistic values of thermal noise ($n_{j} \simeq 2\times 10^7$ corresponding to thermal noise for \SI{305}{\kilo\hertz} mechanical modes at room temperature, $T=\SI{300}{\kelvin}$).
Particularly the drop in the EPR variance and noise reduction factor shows that the addition of the dispersive coupling helps to symmetrise the mechanical steady state.
This observation is further supported by the state purity which remains high for a broad range of coupling-strength ratios $\lambda_2/\lambda_1$.
Finally, the considered mechanical quality factors ranging between $10^{8}$ and $10^{9}$, which allow the EPR variance to drop below the classical bound of unity, have been realised in recent experiments~\cite{Delic2019,Delic2020} or can be reached with moderate improvements to the vacuum;
the strength of the dispersive coupling considered here is feasible in levitated systems as well~\cite{Delic2020a}.

To further confirm our intuition regarding the collective Bogoliubov modes, we investigate the variation of entanglement with the dispersive coupling rates in figure~\ref{fig_3}.
For a fixed value of the dispersive coupling $g_1$, there exists an optimum coupling $g_2$ that simultaneously maximises the logarithmic negativity and purity while minimizing the EPR variance.
When $g_2$ is too small, the cooling of the second Bogoliubov mode, $\propto g_2\lambda_1\beta_2 a_2^\dagger + {\rm H.c.}$, is weak;
large $g_2$, on the other hand, enhances the single-mode squeezing of the first Bogoliubov mode, $\propto g_2\lambda_2\beta_1 a_2 + {\rm H.c.}$, eventually leading to instability.
Moreover, as the value of $g_1$ starts approaching the coherent-scattering coupling (the dotted lines in figure~\ref{fig_3}), the generated entanglement starts to drop since the dispersive coupling begins to damp the quantum correlations generated by the coherent-scattering interaction.
These results thus confirm our understanding and show that there is a broad range of dispersive coupling strengths for which entanglement between the particles can be generated.

\subsection{Floquet analysis of oscillating terms}

So far we have shown the results obtained under the rotating wave approximation.
To provide a more detailed analysis of our proposal under realistic experimental conditions, we now analyse two situations where the rotating wave approximation cannot be applied.
First, we study the effect of counterrotating terms omitted from the Hamiltonian~\eqref{int_Ham} and show that entanglement can still be generated when realistic values for the sideband ratio $\kappa_j/\Omega_1$ and coupling $\lambda_j/\Omega_1$ are included.
Next, we consider a scenario where the two mechanical frequencies are different, $\Omega_1\neq\Omega_2$, and demonstrate that our scheme is capable of generating strong entanglement in this case as well.

To analyse dynamics beyond the rotating wave approximation, we start from the Hamiltonian~\eqref{rfHamil} appended with the dispersive part of the interaction, $H_{\rm disp} = \Delta_2a_2^\dagger a_2+g_1(a_2+a_2^\dagger)(b_1+b_1^\dagger) + g_2(a_2+a_2^\dagger)(b_2+b_2^\dagger)$, and move to the rotating frame with respect to the free oscillations, $H_{0}=\Delta_{1} a_{1}^\dag a_{1}+\Delta_{2} a_{2}^\dag a_{2}+ \Omega_1 b_{1}^\dag b_{1}+\Omega_2 b_{2}^\dag b_{2}$.
Assuming again $\Omega_1 = \Omega_2$, we obtain the full interaction-picture Hamiltonian
\begin{align}
\begin{split}
	H &=-\lambda_1(a_{1}^\dagger b_{1}+b_{1}^\dagger a_{1})-\lambda_2(a_{1}b_{2}+ a_{1}^\dagger b_{2}^\dagger) -\lambda_{1}(a_{1}b_{1}e^{-2i\Omega_1 t}+a_{1}^\dag b_{1}^\dag e^{2i\Omega_1 t}) -\lambda_{2}(a_{1}b_{2}^\dag e^{2i\Omega_1 t}+a_{1}^\dag b_{2}e^{-2i\Omega_1 t})\\
	&\quad + g_1(a_2^\dagger b_1+b_1^\dagger a_2) + g_2(a_2^\dagger b_2 + b_2^\dagger a_2) +g_{1}(a_{2}b_{1}e^{-2i\Omega_1 t}+a_{2}^\dag b_{1}^\dag e^{2i\Omega_1 t}) + g_{2}(a_{2}b_{2}e^{-2i\Omega_1 t}+a_{2}^\dag b_{2}^\dag e^{2i\Omega_1 t}),
\end{split}
\end{align}
from which we express the time-dependent Langevin equations for the quadrature operators
\begin{subequations}
\begin{align}
	\dot{r} &= A(t)r + \xi,\\
	A(t) &= \begin{pmatrix}
		-\frac{1}{2}\kappa_{1} &0&0&0& -\lambda_{1}s(t)&-\lambda_{1}c_-(t) & \lambda_{2}s(t) & \lambda_{2}c_-(t) \\
		0& -\frac{1}{2}\kappa_{1} &0&0& \lambda_{1}c_+(t) & \lambda_{1}s(t) & \lambda_{2}c_+(t) & \lambda_{2}s(t) \\
		0&0& -\frac{1}{2}\kappa_{2} &0&  g_{1}s(t) & g_{1}c_-(t) & g_{2}s(t) & g_{2}c_-(t) \\
		0&0&0& -\frac{1}{2}\kappa_{2} & -g_{1}c_+(t) & -g_{1}s(t) & -g_{2}c_+(t) & -g_{2}s(t) \\
		-\lambda_{1}s(t) & -\lambda_{1}c_-(t) & g_{1}s(t) & g_{1}c_-(t) & -\frac{1}{2}\gamma_{1} &0&0&0 \\
		\lambda_{1}c_+(t) & \lambda_{1}s(t) & -g_{1}c_+(t) & -g_{1}s(t) &0& -\frac{1}{2}\gamma_{1} &0&0 \\
		-\lambda_{2}s(t) & \lambda_{2}c_-(t) & g_{2}s(t) & g_{2}c_-(t) &0&0& -\frac{1}{2}\gamma_{2} &0& \\
		\lambda_{2}c_+(t) & -\lambda_{2}s(t) & -g_{2}c_+(t) & -g_{2}s(t) &0&0&0& -\frac{1}{2}\gamma_{2}
\end{pmatrix},
\end{align}
\end{subequations}
where $c_\pm(t) = 1\pm\cos(2\Omega_1 t)$, $s(t) = \sin(2\Omega_1 t)$.
To find the steady state in this case, we express the drift matrix $A(t)$ in the Floquet space~\cite{Pietikainen2020}.
Using the approach outlined in the appendix, we can thus turn the time-dependent problem into a time-independent one for which the corresponding Lyapunov equation can be solved to obtain the steady state covariance matrix.

\begin{figure}
	\centering
	\includegraphics[width=0.95\linewidth]{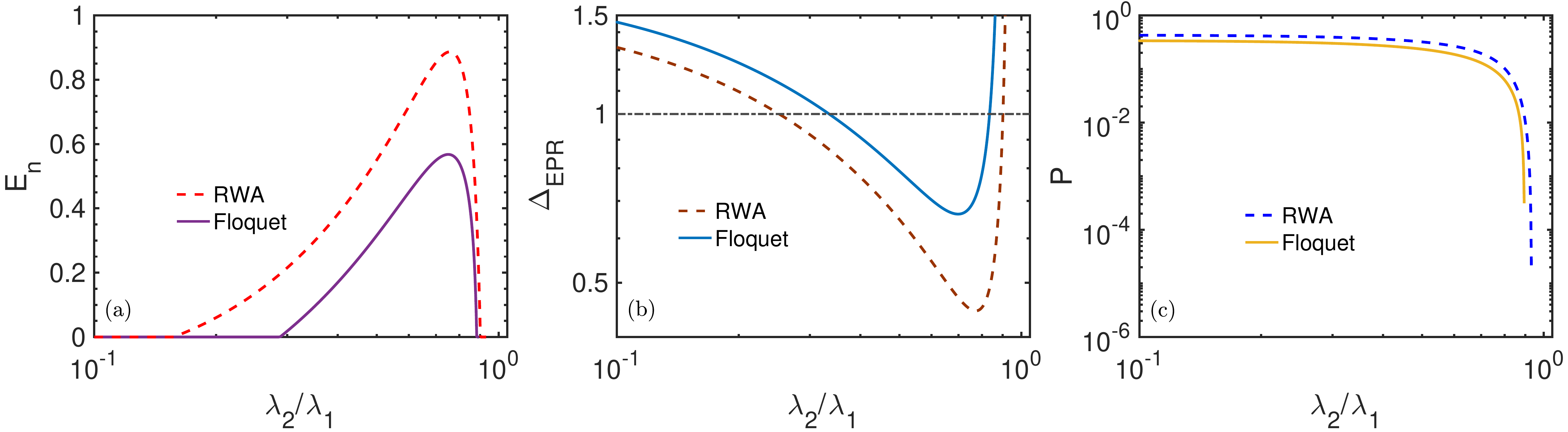}
	\caption{\label{fig_5}
		Comparison of the rotating wave approximation and full dynamics.
		We plot the logarithmic negativity (a), EPR variance (b) and state purity (c) for RWA (dashed) and Floquet--Lyapunov method (solid) for $\lambda_{1}/2\pi=\SI{100}{\kilo\hertz}$, $\kappa_{j}/2\pi=\SI{120}{\kilo\hertz}$, $g_{1}/2\pi=\SI{3}{\kilo\hertz}$, $g_{2}/2\pi=\SI{20}{\kilo\hertz}$, $Q_j = 10^9$, and $n_{j}=2\times10^7$;
		the result for the rotating wave approximation is thus the same as used in figure~\ref{fig:results}.}
\end{figure}

Results of the corresponding simulations are shown in figure~\ref{fig_5}.
The finite sideband ratio ($\kappa_j/\Omega_1\simeq 0.4$ for the experiment in~\cite{Delic2020}) and strong coupling ($\lambda_1/\Omega_1\simeq 0.35$) allow generation of entanglement even when the effect of counterrotating terms is taken into account.
Crucially, the noise due to the counterrotating terms is distributed evenly between the two mechanical modes, allowing the presence of entanglement to be verified from the symmetric EPR variance.

Another situation where the rotating wave approximation cannot be applied arises when the two mechanical frequencies are different, $\Omega_1\neq\Omega_2$, which might be the case for two nanoparticles of different sizes.
In this case, the interactions in the coherent-scattering Hamiltonian $H_{\rm cs}$ can still remain resonant when suitable tweezer frequencies are used as discussed above;
the Bogoliubov mode $\beta_1$ can thus still be efficiently cooled by coherent scattering.
The dispersive coupling with a single driving frequency (set on the red sideband of the second mechanical mode), on the other hand, becomes generally time-dependent and cannot efficiently cool the Bogoliubov mode $\beta_2$.

The most natural choice is then to make the interaction between the modes $a_2$ and $b_2$ resonant as this mechanical mode is---unlike $b_1$---not cooled down by the coherent-scattering interaction.
We thus obtain the dispersive Hamiltonian
\begin{equation}\label{eq:detuning}
H_{\rm{disp}}= g_1 a_2b_1^\dagger e^{i\delta_{12} t} + g_2 a_2b_2^\dagger + {\rm H.c.},
\end{equation}
where $\delta_{12}=\Omega_{1}-\Omega_{2}$ is the detuning between the two mechanical modes and we applied the rotating wave approximation to neglect terms oscillating at $2\Omega_{1,2}$ and $\Omega_1+\Omega_2$.
To solve the resulting time-dependent Lyapunov equation in the steady state, we again express the periodic dynamics in the Floquet space and solve the resulting time-independent version as described in the appendix.
The results of this simulation in figure~\ref{fig_4}(a)--(c) show that the detuning has no observable effect on the generated entanglement, proving that strong quantum correlations can be efficiently prepared also with particles of unequal frequencies.

\begin{figure}
	\centering
	\includegraphics[width=0.95\linewidth]{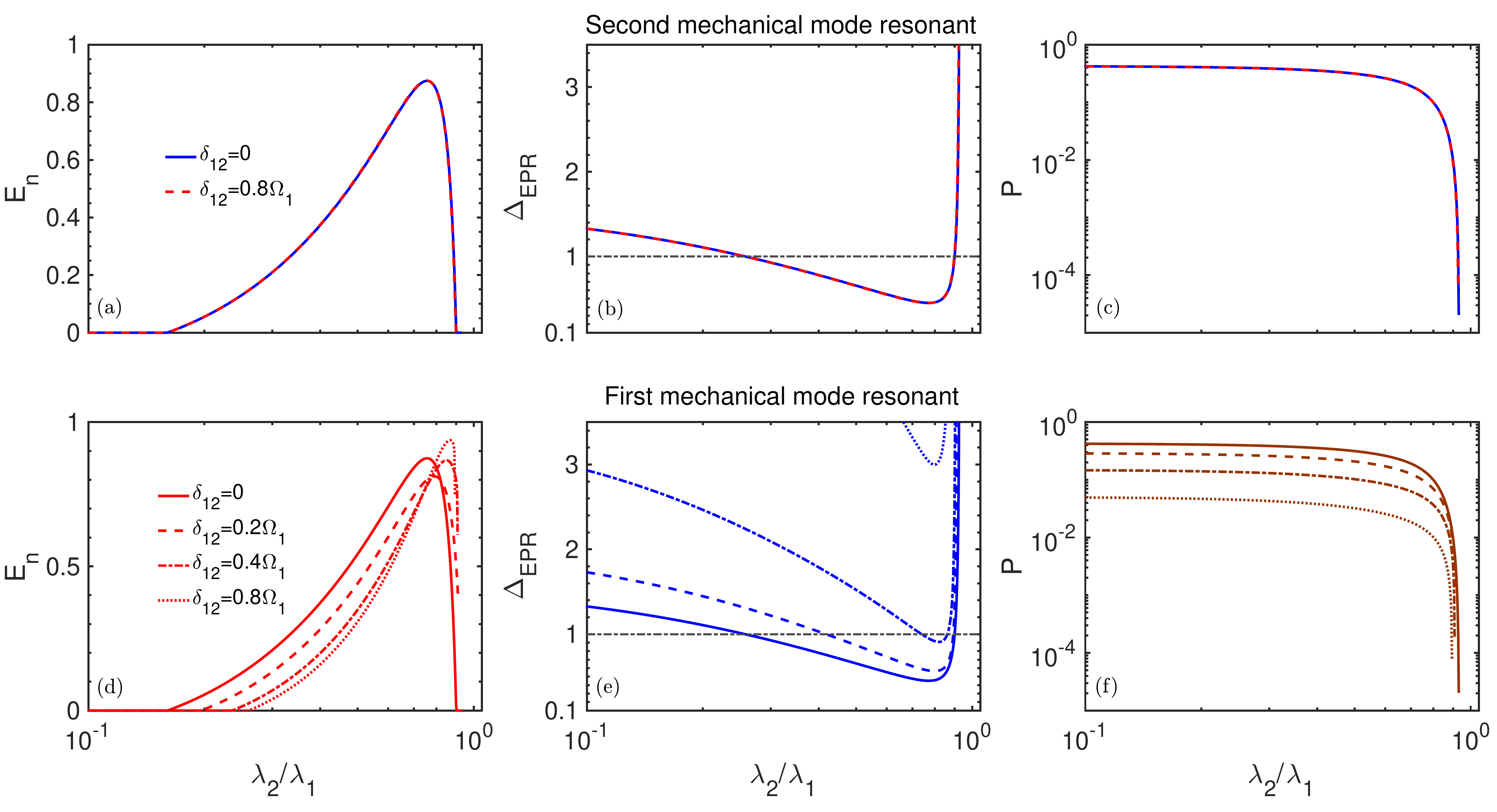}
	\caption{\label{fig_4}
	Entanglement generation with unequal mechanical frequencies.
	(a)--(c) The second mechanical mode being resonant in the dispersive interaction as described by~\eqref{eq:detuning}.
	We again show results for (a) the logarithmic negativity, (b) EPR variance, and (c) purity versus the ratio of coherent scattering coupling strengths $\lambda_{2}/\lambda_{1}$.
	We show results for $\delta_{12}=0$ (solid blue line) and $\delta_{12}/2\pi=\SI{240}{\kilo\hertz}$ (dashed red line).
	(d)--(f) The first mechanical mode resonant in the dispersive interaction as described by~\eqref{eq:detuning2}.
	We plot (d) the logarithmic negativity, (e) EPR variance, and (f) state purity for $\delta_{12}=0$ (solid line), $\delta_{12}/2\pi=\SI{60}{\kilo\hertz}$ (dashed), $\delta_{12}/2\pi=\SI{120}{\kilo\hertz}$ (dot-dashed), and $\delta_{12}/2\pi=\SI{240}{\kilo\hertz}$ (dotted).
	The remaining parameters are the same as in figure~\ref{fig:results}.}
\end{figure}

If, on the other hand, the second cavity field is resonant with the first mechanical mode as described by the Hamiltonian
\begin{equation}\label{eq:detuning2}
	H_{\rm{disp}}= g_1 a_2b_1^\dagger + g_2 a_2b_2^\dagger e^{-i\delta_{12} t} + {\rm H.c.},
\end{equation}
the performance of our scheme worsens which we show in figure~\ref{fig_4}(d)--(f).
The logarithmic negativity of the mechanical steady state remains high even for a large detuning between the two mechanical modes;
however, as the second mechanical mode becomes off-resonant with the cavity field, the state purity decreases and the EPR variance increases.
This behaviour shows that this regime is not well suited for entanglement generation, further supporting our understanding that the dispersive coupling needs to efficiently remove noise from the second mechanical mode $b_2$.

\section{Experimental feasibility}

\subsection{Nanoparticle decoherence}

For our simulations, we worked with parameters close to the recent experimental demonstrations of optomechanical cooling via coherent scattering~\cite{Delic2019,Delic2020}.
The main differences are a slightly improved cavity decay rate (achievable with an improved coating of the cavity mirrors) and smaller thermal decoherence rate of the mechanics (possible by reducing the pressure in the vacuum chamber from \SI{e-6}{\milli\bar} used in Ref.~\cite{Delic2020}).
Note, however, that our model does not assume decoherence merely by gas damping.
In our effective, phase-insensitive decoherence we include a broad range of effects.
The crucial parameter in our simulations is the thermal decoherence rate $\gamma n_j$ which ought to be compared with experimentally available heating rates.
Admittedly, our heating rate $\gamma_j n_j/2\pi\simeq \SI{6}{\kilo\hertz}$ (corresponding to $Q_j = \Omega/\gamma_j = 10^9$) is smaller than the total heating rate reported in Ref.~\cite{Delic2020};
nevertheless, it corresponds to the recoil heating rate reported there.

Moreover, the values of dispersive coupling considered here are also within reach of current experiments.
Coupling approaching $g/\Omega\simeq 0.1$ has recently been demonstrated, albeit with a smaller mechanical frequency (of about \SI{163}{\kilo\hertz}) and coupling rate (of about \SI{14.4}{\kilo\hertz})~\cite{Delic2020a}.
Given the broad range of experimental parameters over which nonclassical two-mode squeezed states can be prepared (see figure~\ref{fig_3}), our proposed scheme is feasible.

Importantly, decoherence processes introduced by the cavity drive do not pose a limitation in our proposal.
First, photon recoil, proportional to the light intensity, is much weaker than the recoil caused by the trapping beam.
The intensity of the tweezer is about \SI{8.2e7}{\watt\per\centi\metre^2} (for power $P = \SI{0.4}{\watt}$ and waist $w = \SI{0.7}{\micro\metre}$~\cite{Delic2020}) whereas the cavity-field intensity is merely \SI{1.9e6}{\watt\per\centi\metre^2} ($P = \SI{30}{\watt}$, $w=\SI{40}{\micro\metre}$; this value corresponds to $n_{\rm phot} = 10^{10}$ photons needed to reach coupling of \SI{30}{\kilo\hertz} from single-photon coupling of \SI{0.3}{\hertz}~\cite{Delic2020a}).
The heating rate associated with the cavity photon recoil can therefore be expected to be of the order of \SI{100}{\hertz}.

A bigger problem associated with the cavity drive is the laser phase noise;
the corresponding heating rate can be expressed as
\[
	\Gamma_{\rm phase} = \frac{4g^2 n_{\rm phot}}{\kappa^2} S_{\dot{\phi}}(\Omega),
\]
where $S_{\dot{\phi}}(\Omega)$ is the frequency noise spectrum.
With typical values of \SI{0.1}{\hertz^2\per\hertz}, we get a heating rate of the order of \SI{100}{\mega\hertz}.
The laser phase noise is, however, only a technical limitation and can be further reduced with the help of a filtering cavity~\cite{Hald2005};
to obtain heating rate comparable to the photon recoil from the tweezer, a filtering cavity with a linewidth of about \SI{40}{\kilo\hertz} is sufficient.

\subsection{Verification of entanglement}

The generated entanglement can be most easily detected via the EPR variance~\eqref{eq:EPR} which requires only estimating the variance of the sum of the mechanical positions, $\Delta(x_1+x_2)$, and of the difference of mechanical momenta, $\Delta(p_1-p_2)$, and not tomography of the full covariance matrix.
The best strategy in terms of measurement precision would employ a third cavity mode (such that the above described interactions still actively stabilise the entangled state) using backaction evading measurements of a single joint mechanical quadrature~\cite{Clerk2008}.
Such a measurement allows, in principle, sensitivity below the standard quantum limit as measurement backaction affects the two orthogonal collective quadratures $x_1-x_2$, $p_1+p_2$;
since the two measured quadratures commute, they can be measured simultaneously (or subsequently) with shot-noise limited precision.

The measurement requires both particles to be coupled to the readout cavity mode with the same rate.
This arrangement ensures that the collective mechanical quadrature coupled to the cavity field contains both particles with the same weight.
As it might be difficult to achieve such coupling via radiation pressure interaction (which would introduce additional photon recoil and laser phase noise), the readout could be performed via coherent scattering as well.
Using a pair of weak additional tweezer beams for each particle---one on the blue sideband, one on the red---would lead to scattering of both Stokes and anti-Stokes photons into the cavity, analogous to two-tone driving in conventional optomechanics~\cite{Clerk2008}.
Such a setting would also have the advantage that, for each particle, the mechanical quadrature coupled to the cavity field would be set by the relative phase between the two tweezers;
the measured quadratures for the two particles can thus be set independently.
This readout therefore allows also measurement of the orthogonal collective quadratures $x_1 - x_2$, $p_1+p_2$, as well as single-particle quadratures, and thus enables full quantum tomography of the mechanical state which can be used to estimate the noise reduction factor~\eqref{deln} or the logarithmic negativity.

\section{Conclusions}
In summary, we presented a scheme that generates entanglement between two levitated nanoparticles deterministically in the steady state.
Coherent scattering of photons from tweezers into an empty cavity mode can be used for two-mode squeezing of the particle motion when suitable detunings between the tweezers and the cavity are used.
We showed how coherent scattering from the tweezers into an empty cavity mode, together with a second, dispersively coupled cavity mode, can be used to cool two mechanical Bogoliubov modes, creating a two-mode squeezed stationary state.
Moreover, the state has sufficiently high purity and strong phonon--phonon correlations to allow further applications.
Crucially, our work demonstrates that it is not necessary to engineer exact cooling dynamics for both Bogoliubov modes which would require complicated multitone driving schemes;
instead, it is sufficient to create such dynamics approximately using coherent scattering and dispersive optomechanical interaction with a single driving tone.

With levitated optomechanics entering the quantum regime, our proposal presents a viable approach towards creating complex nonclassical states of massive objects at room temperature.
Our results show an attractive way towards engineering quantum dynamics and states of collective motion of multiple particles with minimal resources with possible extensions to the preparation of similar states across macroscopic distances or with more than two particles.
In classical levitated optomechanics, multiparticle effects have already been observed~\cite{Arita2018};
extending the dynamics of multiple particles to the quantum regime presents an important steps forward for both basic and applied science, allowing new tests of decoherence models and schemes to measure weak forces or fields.

\begin{acknowledgments}
	We would like to thank Uro\v{s} Deli\'{c} for useful discussions regarding experimental feasibility of our proposal and decoherence mechanisms for the particle motion.
	We gratefully acknowledge support by the project 20-16577S of the Czech Science Foundation, project CZ.02.1.01/ 0.0/0.0/16\textunderscore{}026/0008460 of MEYS \v{C}R, and European Union's Horizon 2020 (2014--2020) research and innovation framework programme under Grant Agreement No. 731473 (project 8C18003 TheBlinQC). Project TheBlinQC has received funding from the Quant\-ERA ERA-NET Cofund in Quantum Technologies implemented within the European Union's Horizon 2020 Programme.
\end{acknowledgments}

\appendix

\section{Lyapunov equation for the covariance matrix of the system}{\label{Lyap}}

To find the steady state of the optomechanical system, we start from the Langevin equations in the matrix form
\begin{subequations}
\begin{align}
	\dot{{r}} &= {Ar}+{\xi},\\
	{r} &= (X_1,Y_1,X_2,Y_2,x_1,p_1,x_2,p_2)^T,\\
	{\xi} &= (\sqrt{\kappa_1}X_{1,{\rm in}},\sqrt{\kappa_1}Y_{1,{\rm in}}, \sqrt{\kappa_2}X_{2,{\rm in}},\sqrt{\kappa_2}Y_{2,{\rm in}}, \sqrt{\gamma_1}x_{1,{\rm in}},\sqrt{\gamma_1}p_{1,{\rm in}}, \sqrt{\gamma_2}x_{2,{\rm in}},\sqrt{\gamma_2}p_{2,{\rm in}})^T \label{eq:SMnoise}\\
	{A} &= \begin{pmatrix}
		-\frac{1}{2}\kappa_{1} &0& 0 &0&0& -\lambda_{1} & 0& \lambda_{2}\\
		0& -\frac{1}{2}\kappa_{1} &0&0& \lambda_{1} & 0& \lambda_{2}  & 0\\
		0& 0& -\frac{1}{2}\kappa_{2}& 0& 0& g_{1}& 0& g_{2}\\
		0& 0&0& -\frac{1}{2}\kappa_{2}& -g_{1}&0& -g_{2}&0\\
		0 & -\lambda_{1} &0 &g_{1}& -\frac{1}{2}\gamma_{{1}} & 0 &   0 & 0 \\
		\lambda_{1} & 0& -g_{1} &0& 0&  -\frac{1}{2}\gamma_{{1}}& 0 & 0 \\
		0&\lambda_{2} & 0 &g_{2}& 0&0 & -\frac{1}{2}\gamma_{{2}} & 0\\
		\lambda_{2} & 0&-g_{2} & 0 & 0 &0& 0 & -\frac{1}{2}\gamma_{2}\\
	\end{pmatrix}.
\end{align}
\end{subequations}
The steady state of the system can now be expressed in terms of the covariance matrix which obeys the Lyapunov equation
\begin{equation}
	{AV}+{VA}^{T}+ {N}=0,
\end{equation}
where ${V}$ with elements $V_{jk}=\langle r_{j}r_{k} +r_{k}r_{j} \rangle-2\langle r_{j}\rangle \langle r_{j}\rangle$ is the covariance matrix of the system's Wigner function and
\begin{equation}\label{eq:SMdiff}
	{N} = \langle {\xi}(t){\xi}^T(t)\rangle
	= {\rm diag}[\kappa_{1}, \kappa_{1},\kappa_{2},\kappa_{2}, \gamma_{{1}}(2n_{1}+1), \gamma_{{1}}(2n_{1}+1), \gamma_{{2}}(2n_{2}+1),\gamma_{2}(2n_{2}+1)]
\end{equation}
is the diffusion matrix.
The properties of the mechanical state (the logarithmic negativity, EPR variance, noise reduction factor, and state purity), are encoded in the lower right $4\times 4$ block of the covariance matrix ${V}$.
To express the noise reduction factor from the covariance matrix, we use the identity (we use $\langle r_i\rangle = 0$ here and in the following since we have no linear terms in the Hamiltonian)
\begin{equation}
	\langle \{r_{i},\{r_{j},\{ r_{k},r_{l} \}\}\}\rangle=2\left( V_{ij}V_{kl}+V_{ik} V_{jl}+V_{il}V_{jk}\right),
\end{equation}
where $\{\cdot,\cdot\}$ denotes the anticommutator.
This formula can be derived from a comparison between the fourth derivative of a general and a Gaussian characteristic function~\cite{Cernotik2017}.
A straightforward calculation reveals that the variance of the phonon-number difference can be expressed as (V now denotes only the mechanical part of the covariance matrix)
\begin{equation}
	\langle(n_1-n_2)^2\rangle - \langle n_1-n_2\rangle^2 = \frac{1}{8}\left( V_{11}^2+V_{22}^2+V_{33}^2+V_{44}^2\right)+\frac{1}{4}\left( V_{12}^2+V_{34}^2-V_{13}^2-V_{14}^2-V_{23}^2-V_{24}^2 \right)-\frac{1}{2}
\end{equation}
while the mean phonon number is given by
\begin{equation}
	\langle n_1\rangle = \frac{1}{4}(V_{11}+V_{22}-2), \qquad \langle n_2\rangle = \frac{1}{4}(V_{33}+V_{44}-2).
\end{equation}

The Lyapunov equation can also be used to find the steady state of a system with periodic Hamiltonian;
in this case, however, it must be expressed in the Floquet space~\cite{Pietikainen2020}.
To include the effect of the counterrotating terms neglected above, we start from the full equations of motion for the quadrature operators
\begin{subequations}
\begin{align}
	\dot{r} &= A(t)r + \xi,\\
	A(t) &= \begin{pmatrix}
		-\frac{1}{2}\kappa_{1} &0&0&0& -\lambda_{1}s(t)&-\lambda_{1}c_-(t) & \lambda_{2}s(t) & \lambda_{2}c_-(t) \\
		0& -\frac{1}{2}\kappa_{1} &0&0& \lambda_{1}c_+(t) & \lambda_{1}s(t) & \lambda_{2}c_+(t) & \lambda_{2}s(t) \\
		0&0& -\frac{1}{2}\kappa_{2} &0&  g_{1}s(t) & g_{1}c_-(t) & g_{2}s(t) & g_{2}c_-(t) \\
		0&0&0& -\frac{1}{2}\kappa_{2} & -g_{1}c_+(t) & -g_{1}s(t) & -g_{2}c_+(t) & -g_{2}s(t) \\
		-\lambda_{1}s(t) & -\lambda_{1}c_-(t) & g_{1}s(t) & g_{1}c_-(t) & -\frac{1}{2}\gamma_{1} &0&0&0 \\
		\lambda_{1}c_+(t) & \lambda_{1}s(t) & -g_{1}c_+(t) & -g_{1}s(t) &0& -\frac{1}{2}\gamma_{1} &0&0 \\
		-\lambda_{2}s(t) & \lambda_{2}c_-(t) & g_{2}s(t) & g_{2}c_-(t) &0&0& -\frac{1}{2}\gamma_{2} &0& \\
		\lambda_{2}c_+(t) & -\lambda_{2}s(t) & -g_{2}c_+(t) & -g_{2}s(t) &0&0&0& -\frac{1}{2}\gamma_{2}
\end{pmatrix},
\end{align}
\end{subequations}
where $c_\pm(t) = 1\pm\cos(2\Omega_1 t)$, $s(t) = \sin(2\Omega_1 t)$.
Next, we express the drift matrix $A(t)$ in terms of its Fourier components defined via the expression
\begin{equation}
	A(t) = A^{(0)} + \sqrt{2}\sum_{n=1}^\infty [A_c^{(n)}\cos(2n\Omega_1 t) + A_s^{(n)}\sin(2n\Omega_1 t)],
\end{equation}
where each of the matrices $A^{(0)},A_{c,s}^{(n)}$ is time independent.
We can now formulate the Lyapunov equation in the Floquet-space,
\begin{equation}
	A_{F}V_{F}+V_{F}A_{F}^T+N_{F} = 0,
\end{equation}
where we introduced the Floquet-space drift matrix
\begin{equation}
	A_{F} =\begin{pmatrix}
		A^{(0)} & A_{c}^{(1)} & A_{s}^{(1)} & 0 & 0 & \ldots \\
		A_{c}^{(1)} & A^{(0)} & -2\Omega_1 I & \frac{1}{\sqrt{2}}A_{c}^{(1)} & \frac{1}{\sqrt{2}}A_{s}^{(1)}\\
		A_{s}^{(1)} & 2\Omega_1 I & A^{(0)} & -\frac{1}{\sqrt{2}}A_{s}^{(1)} & \frac{1}{\sqrt{2}}A_{c}^{(1)} \\
		0 & \frac{1}{\sqrt{2}}A_{c}^{(1)} & -A_{s}^{(1)} & A^{(0)}& -4\Omega_1 I\\
		0 & \frac{1}{\sqrt{2}}A_{s}^{(1)} & \frac{1}{\sqrt{2}}A_{c}^{(1)} & 4\Omega_1 I & A^{(0)} \\
		\vdots &&&&& \ddots
	\end{pmatrix}
\end{equation}
with $I$ denoting the $8\times 8$ identity matrix and the Floquet-space diffusion matrix $N_{F}={\rm diag}( N, N, N, N, N, \ldots)$ and $N$ defined in~\eqref{eq:SMdiff}.
The solution is now contained in the zeroth frequency block of the covariance matrix $V_F$ (in its upper left $8\times 8$ corner);
the mechanical covariance matrix is obtained from this component the same way as for the previous case with the rotating wave approximation.

The Floquet--Lyapunov approach can also be applied to the case of unequal mechanical frequencies.
Starting from the rotating-frame Hamiltonian $H_{\rm cs} + H_{\rm disp}$, where $H_{\rm disp}$ is given in~\eqref{eq:detuning}, we obtain the Langevin equations
\begin{subequations}
\begin{align}
	\dot{r} &= A(t)r + \xi,\\
	A(t) &= \begin{pmatrix}
		-\frac{1}{2}\kappa_{1} &0& 0 & 0 & 0 &-\lambda_{1} & 0& \lambda_{2}\\
		0& -\frac{1}{2}\kappa_{1} &0&0 & \lambda_{1}& 0& \lambda_{2}& 0\\
		0&0&-\frac{1}{2}\kappa_{2} &0&  -g_{1}\sin(\delta_{12}t)&g_{1}\cos(\delta_{12}t)& 0&g_{2} \\
		0&0&0&-\frac{1}{2}\kappa_{2} &  -g_{1} \cos(\delta_{12}t)& -g_{1}\sin(\delta_{12}t)& -g_{2}& 0\\
		0&-\lambda_{1} & g_{1} \sin(\delta_{12}t)& g_{1} \cos(\delta_{12}t) & -\frac{1}{2}\gamma_{1} &0& 0 & 0\\
		\lambda_{1}& 0& -g_{1} \cos(\delta_{12}t)&g_{1} \sin(\delta_{12}t) & 0& -\frac{1}{2}\gamma_{1} &0&0\\
		0&\lambda_{2} & 0& g_{2}& 0&0&-\frac{1}{2}\gamma_{2} &0&\\
		\lambda_{2}& 0& -g_{2}& 0& 0&0&0&-\frac{1}{2}\gamma_{2}
	\end{pmatrix}, \label{eq:SMdrift2res}
\end{align}
\end{subequations}
where the vector of input noise operators is the same as before and is given by~\eqref{eq:SMnoise}.
The rotating wave approximation cannot be applied to the time-dependent part of the evolution since $\delta_{12} = \Omega_1-\Omega_2$ is not necessarily larger than the cavity decay and optomechanical coupling rates.
We can, however, expand the drift matrix into its Fourier components,
\begin{equation}
	A(t) = A^{(0)} + \sqrt{2}\sum_{n=1}^\infty [A_c^{(n)}\cos(n\delta_{12} t) + A_s^{(n)}\sin(n\delta_{12} t)],
\end{equation}
and solve for the steady state in the Floquet space using the same approach as for the counterrotating terms above;
the results are shown in figure~\ref{fig_4}(a)--(c) of the main text.

When the second cavity mode $a_2$ is driven on the red sideband of the first mechanical mode $b_1$ as described by~\eqref{eq:detuning2}, the drift matrix becomes
\begin{equation}\label{eq:SMdrift1}
	A(t) = \begin{pmatrix}
		-\frac{1}{2}\kappa_{1} &0& 0 & 0 & 0 &-\lambda_{1} & 0& \lambda_{2}\\
 		0& -\frac{1}{2}\kappa_{1} &0&0 & \lambda_{1}& 0& \lambda_{2}& 0\\
		0&0&-\frac{1}{2}\kappa_{2} &0&  0&g_{1}& g_{2}\sin(\delta_{12}t)& g_{2}\cos(\delta_{12}t)\\
		0&0&0&-\frac{1}{2}\kappa_{2} &  -g_{1}& 0& -g_{2}\cos(\delta_{12}t)& g_{2}\sin(\delta_{12}t)\\
		0&-\lambda_{1} & 0& g_{1} & -\frac{1}{2}\gamma_{1} &0& 0 & 0\\
		\lambda_{1}& 0& -g_{1}&0 & 0& -\frac{1}{2}\gamma_{1} &0&0\\
		0&\lambda_{2} & -g_{2}\sin(\delta_{12}t)& g_{2}\cos(\delta_{12}t)& 0&0&-\frac{1}{2}\gamma_{2} &0&\\
		\lambda_{2}& 0& -g_{2}\cos(\delta_{12}t)& -g_{2}\sin(\delta_{12}t)& 0&0&0&-\frac{1}{2}\gamma_{2}
\end{pmatrix}.
\end{equation}
The Floquet--Lyapunov approach is applied in the same way and leads to the results shown in figure~\ref{fig_4}(d)--(f).

%

\end{document}